# Coupling Optimization using Design Structure Matrices and Genetic Algorithm

Sébastien Dubé
Samares-Engineering, Blagnac
sebastien.dube@samares-engineering.com

Mirna Ojeda
Samares-Engineering, Blagnac
mirna.ojeda@samares-engineering.com

Jean-Marie Gauthier
IRT Saint Exupery Toulouse
jean-marie.gauthier@irt-saintexupery.com

**Abstract**    This article seeks to contribute to a nuanced understanding of the integration of Design Structure Matrix (DSM) and genetic algorithms in the context of Complex Systems modelling described within Model-Based System Engineering approach. By examining coupling minimization as a critical aspect of advanced systems engineering practices, we aim to provide a scholarly exploration, blending theoretical insights with practical applications. The objective is to equip systems architects with analytical tools integrated within their Model Based Systems Engineering (MBSE) environment for exploring the design space of component interactions, facilitating the identification of optimal system architectures.

## I. Context and Motivations

Within the domain of Systems Engineering and Model-Based Systems Engineering (MBSE), the efficient management of complex systems necessitates a systematic approach to manage and refine the interactions among their functional components. A fundamental principle within Systems Engineering, emphasized by standards ISO15288:2023 [1] and its associated Systems Engineering Handbook [2], is the imperative to reduce coupling between subsystems for effective management of product complexity. The crux of coupling minimization involves the disentanglement of interdependencies between components.

In most industrial processes, Systems Engineering discipline involves applying Functional Analysis [3] approach and elicit the stakeholder needs using decomposition of system needs into functions and their associated data flow. Then, systems engineering methods propose to distribute these functions over the systems / components implementing the architectural solution. This requires an intermediate "logical architecture" that abstracts function closer to their eventual physical implementations, primarily guided by functional dependencies to ensure coherent integration.

In 2015 a DSM/N² matrix technique [4], has been used for extracting coupling metrics from Functional Architectures, with the capabilities of a genetic algorithm to minimize coupling among logical components. Our investigation addresses this task by employing Coupling Matrices, to assess and reorganize functional dependencies between logical components. The specific emphasis is on coupling minimization as an isolated criterion, accounting for factors such as allocation constraints and timing requirements. Consequently, our method initiates with the identification of functional dependencies, utilizing them to quantify coupling. Then, we propose an optimized allocation of functions to components.

In the subsequent sections, we present our approach consisting in coupling optimization through the application of a genetic algorithm. Following this, we present and discuss the results derived from the application of our approach on two case studies. Finally, we review the relevant literature concerning the application of DSM/N² diagrams in the context of Model-Based Systems Engineering (MBSE).

## II. Background on Genetic Algorithm, N2 Matrices and Capella

In this section, we provide an overview of the integration of the DSM/N² matrix technique and genetic algorithm for coupling optimization.

### a. Optimization of architecture using DSM/N2 Matrices

The N² (N-squared) Matrices, also known as a Coupling Matrix or Design Structure Matrix (DSM), is a graphical representation used in Systems Engineering to analyse and visualize the relationships and dependencies between different components within a system. The primary objective of an N² Matrix is to assess and quantify the coupling or interdependencies between various elements, such as functions, subsystems, or components. The interdependencies within this matrix are defined by functional exchanges connecting two functions, with a binary representation: '1' indicating dependence and '0' signifying independence. Then, the method intends to propose a group of functions (Modules) by successively create groups where the number of interfaces between groups is minimized as illustrated in Figure 1. The sum of interactions outside the module constitutes total interactions.

To perform such grouping, at each iteration, method intends to compute the resulting "coupling" value, which is used to evaluate the resulting complexity of the proposed architecture.

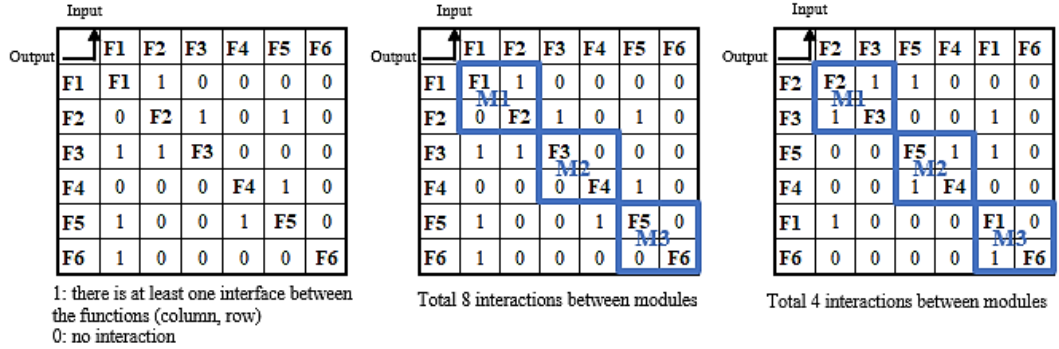

Figure 1. Illustration of the modularization concept and result after minimization of coupling

Using this matrix, we can calculate a coupling value pertaining to the interfaces specified between the logical components, deduced from the interfaces established among functions allocated to these components. The coupling value serves as an assessment of the complexity of coupling between logical components, derived from a formula based on software coupling metrics [5]. For this study, we used the Equation 1 to calculate the coupling value of each individual logical component. Then, the Equation 2 is used to calculate the coupling value of the complete architecture.

$$Coupling(C_{M_k}) = 1 - \frac{1}{d_i + 2 \cdot c_i + d_o + 2 \cdot c_o + \omega + r}$$

Equation 1 - Coupling Value of a Logical Component

$$CouplingValue(C_v) = \sum_{k=1}^{n}[C_{M_k}]$$

Equation 2 - Coupling Value of the Complete Logical Architecture

Where $M_k$ is the logical component under consideration, $d_i$ is the number of input data parameters, $c_i$ is the number of input control parameters, $d_o$ is the number of output data parameters, $c_o$ is the number of output control parameters, $\omega$ is the number of modules called (fan-out), and $r$ is the number of calling the module under consideration (fan-in).

b. Genetic Algorithm

Genetic algorithms, aim to explore the solution space of a given problem to meet predefined criteria. As depicted by the Figure 2, the algorithm initiates by randomly generating an initial population, with each subject representing a potential set of function allocations (1). Subsequently, a fitness function (2), in our case, the coupling equation, assesses each subject, assigning a value or rank that reflects its proximity to the optimal solution. Subjects too distant from the desired solution are then eliminated (3).

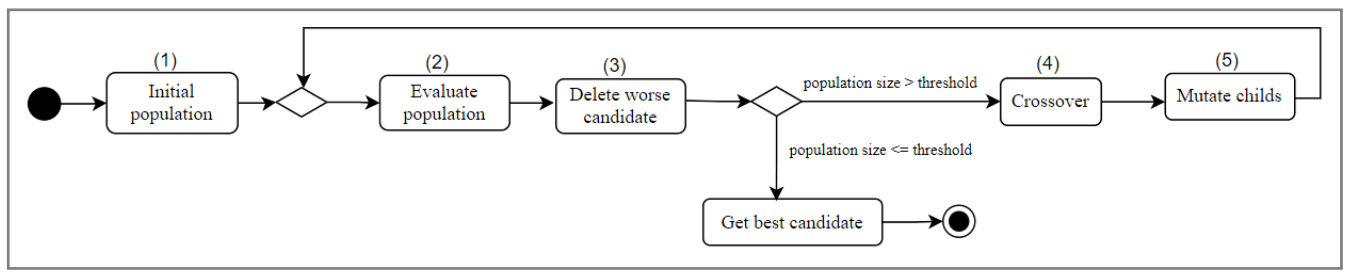

Figure 2. Genetic algorithm processes

The algorithm assesses the remaining subjects, and if the population size is within a specified threshold, it returns the best solution among them; otherwise, it proceeds. The pivotal biomimicry aspect of the genetic algorithm unfolds as the remaining subjects undergo crossover, exchanging genes to generate new subjects (4). These newly created offspring undergo mutation (5), where a portion of their characteristics undergoes random changes. Crossover and mutation serve the purpose of preventing convergence to local optima by dispersing new subjects throughout the solution space.

Genetic algorithms offer configurability through a set of parameters:
- Initial population size, a crucial parameter ensuring sufficient coverage of the solution space at the beginning.
- Max generation number, a parameter determining the algorithm's termination, even as the population grows.
- Percentage of survivors, indicating the proportion of the least fit subjects to be eliminated.

- Percentage of parents, denoting the proportion of subjects participating in crossover.
- Percentage of children to mutate, is the proportion of new subjects subjected to mutation after crossover.
- Percentage of genes to mutate, indicating the proportion of genes to be mutated for each new subject.

As the algorithm is iterative, the conditions for exiting the loop are defined as the number of chromosomes being less than 3 and reaching the maximum number of generations.

To implement programmatically the genetic algorithm, we have been established: the chromosome encoding is the logical component index, the index of chromosomes is the index of each function. The encoded chromosome is represented in 1-dimensional array format where each chromosome has an identifier. An illustration of the GA translation in code becoming in a single gene is shown in the Figure 3. In this example, there are 4 components and 6 functions, as illustration of the allocation mechanism: the function index 1 and 4 is allocated to the component index 2.

| Chromosomes | 1 | 2 | 3 | 4 | 2 | 3 |
|---|---|---|---|---|---|---|
| Function identifier | 0 | 1 | 2 | 3 | 4 | 5 |

*Figure 3. Encoding chromosomes*

### c. Capella and Arcadia

The ARCADIA method [6] defines systems engineering concepts from needs analysis to architectural solution definition, we propose in Figure 4 the highlight of involved concepts in our current proposal:

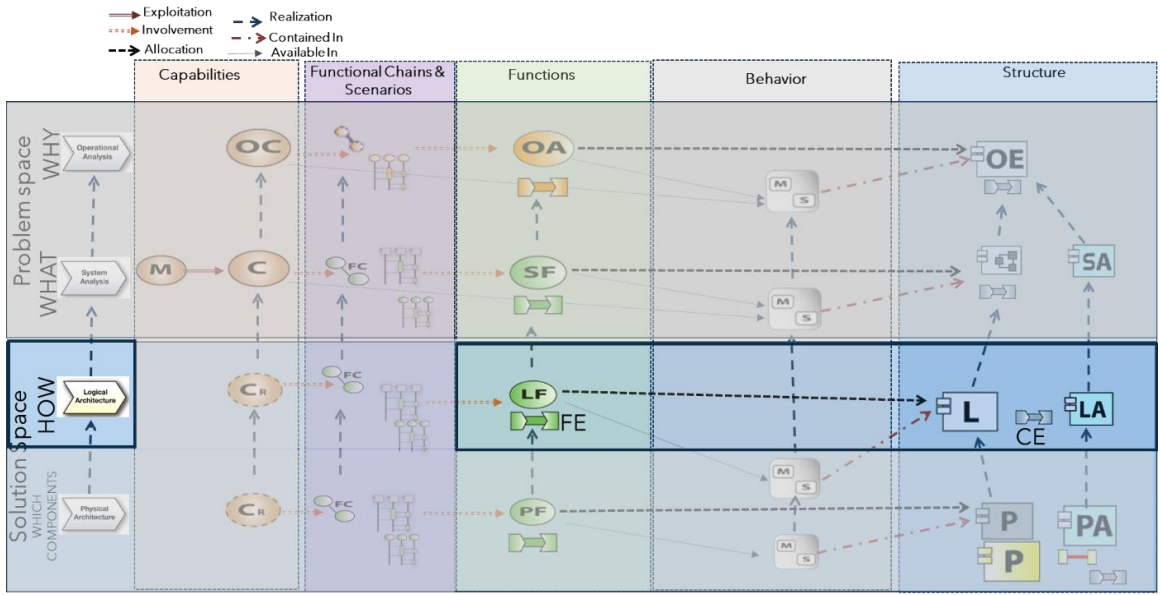

*Figure 4. Overview of ARCADIA concepts and emphasis of involved concepts within our proposal*

So, we consider in our current implementation the Logical Functions (*LF*) allocation to the Logical Components (*L*). The information exchange between functions is modelled using the concept of Functional Exchanges (*FE*). The Functional Exchanges (*FE*) are allocated to Component Exchanges (*CE*) when these exchanges cross the boundaries of a single component. The Component Exchanges (*CE*) represent the interfaces between components and the objective of our proposal consist of to minimize the number of cross-exchanges between components (and minimize interfaces coupling between systems).

An illustration of simplified logical architecture diagram in Capella is illustrated in Figure 5:

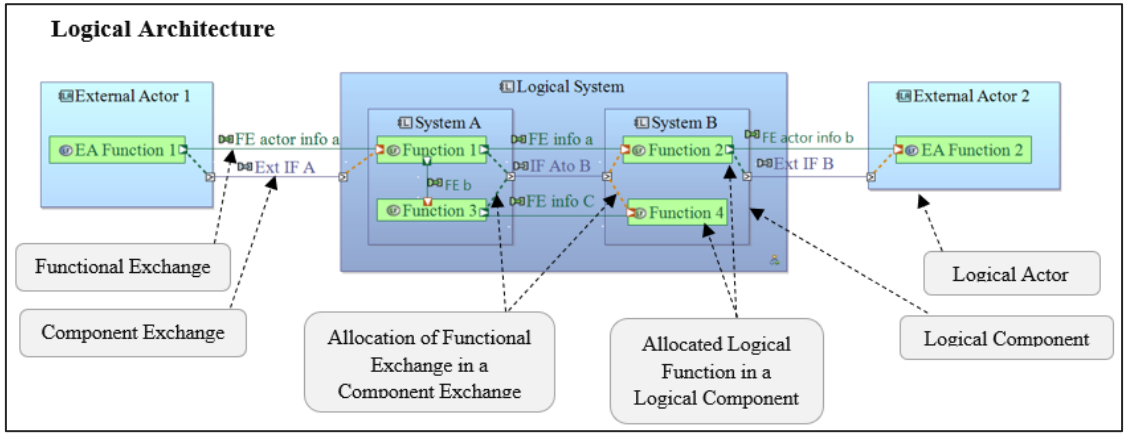

*Figure 5. Simplified Logical Architecture Diagram example and associated concepts*

## III. Implementation and Case Study

In this section, we first present the implementation of our DSM Genetic Algorithm solution. Second, we present the application of our methodology through a case study. Finally, we present and discuss the obtained results concerning the efficiency and the implications of the proposed approach in the domain of Model-Based Systems Engineering.

### a. Implementation in Python and Capella

The implementation of our DSM-based optimization has been integrated within different MBSE approaches and tools. This work relies on our previous work in [7] implemented within Cameo tool. This paper intends to focus on the integration of this technique within Capella MBSE tool [7] and offer to systems engineer facilities to evaluate the more optimized way to distribute functions in their logical architecture in order to minimize interfaces between the defined systems. This automation uses the Python4Capella extension [8], which provides a Python API to interface directly with the Capella model. The algorithm overview is presented below in Figure 6.

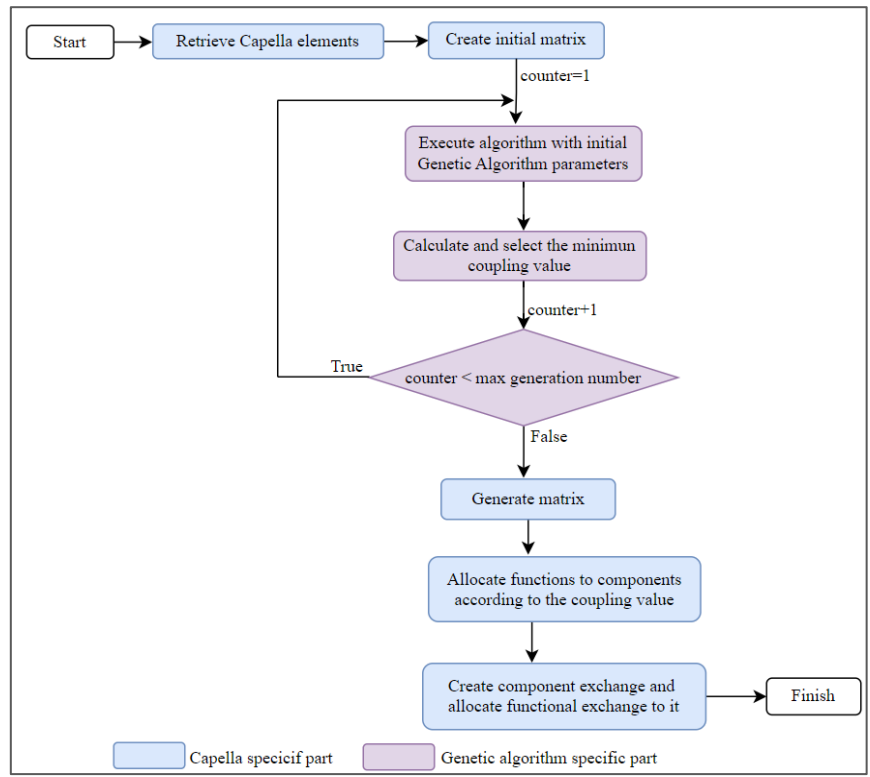

*Figure 6. Algorithm structure and adaptation to Capella*

Our process begins by employing Python4Capella's capabilities to extract the existing logical architecture. Then, using Python scripts, we generate the DSM to visualize and analyse the interdependencies and extract in excel format the resulting DSM Matrix.

The binary DSM is built based on the presence or absence of functional exchanges between functions. Once the DSM is established, the genetic algorithm, configured with initial parameters begins its optimization cycle. The initial parameters are based on our previous work in [9]. After identifying the allocation of functions to components, our tool automates the allocation process within the Capella model. It ensures that the allocations respect pre-defined constraints, such as pre-allocated functions as show in the Figure 7. available functions and are consistent with the lowest coupling values determined by the genetic algorithm.

The final step involves scripting Python4Capella to interpret the updated DSM and translate component interdependencies into Component Exchanges, in which the Functional Exchanges are allocated. This transformation is executed while filtering out relationships within the same component to maintain a focus on inter-component interactions.

The main differences between our current Capella implementations in regards to our initial implementation in Cameo/SysML environment are:

- In Capella, the algorithm considers the Functional Exchanges (*FE*) between all functions. In Capella/ARCADIA, some functions are assigned to the actors defined as external entities in interaction within the system of interest.

- In our initial implementation in Cameo tool, we have used in our functional architecture specialization for exchanges between functions by distinguishing **Flow of Information**, **Energy** or **Matter**. Then, we have created categories of Information Flow: **Data**, **Event**, **Enable/Disable** where "Events" are considered differently in the dependency flow. In our current Capella implementation, we have not yet considered the concept of "Exchanged Items" which may be used to distinguish Events and Shared Data.

In order to get the best solution under the parameters given, we observed that the number of functions to allocate is correlated with the initial population size, the maximum generation number and the gene mutation percentage. Indeed, the higher the number of functions to allocate, the bigger is the space to explore. Therefore, it might-be long process to get a suitable solution sufficiently optimized, especially when the number of candidate solutions grows. So, in further work, we plan to explore the way to minimize the time to obtain a coupling value minimized for large architectures, how to configure automatically initial parameters according to the complexity of architectures, and to study other optimization algorithms to find the best solution.

### b. Application to AIDA Case Study

To assess the relevance of the proposed approach, we applied the optimization process to some case studies. The approach has been applied on an Aircraft Inspection by Drone Assistant system [10]. The AIDA system is a Remotely Piloted Aircraft System (RPAS) that it is composed of a quadcopter drone, a ground station system and a remote control.

#### i. Functions allocation constraints

In the reality of our system there are constraints on how functions should be allocated, influenced by factors such as safety or subcontracting, which is taken in our approach. Figure 7 illustrate this initial functions to components "allocations" constraints that should be respected while proposing a final and complete logical architecture. The figure shows that 15 of 23 logical functions have been allocated among 12 logical components.

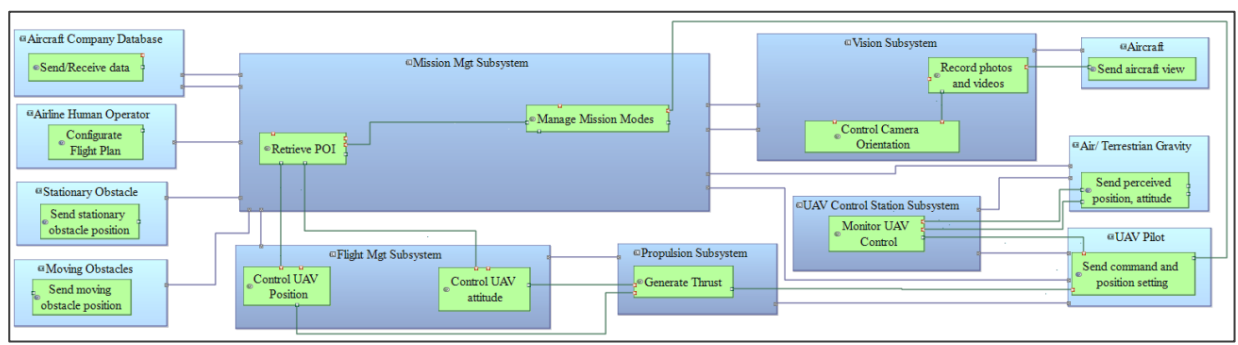

*Figure 7. Architecture before algorithm execution*

In order to get a suitable solution under the parameters given, we observed that the number of functions to allocate is correlated with the initial population size, the maximum generation number and the gene mutation percentage. Indeed, the higher the number of functions to allocate, the bigger is the space to explore. Therefore, it might-be long process to get close to a coupling value minimized, especially when the number of candidate solutions grows.

### ii. Capture of relationships between logical components

The approach produces an initial matrix representing functions and their relationships, as depicted in Figure 8 (derived from the functional architecture and associated functional exchanges). Figure 9 illustrates the resulting matrix after the algorithm's execution, showcasing the functions allocated to 3 out of the 12 components in our use case. These matrices are generated in Excel format by our proposed implementation.

*Figure 8. Initial matrix generated*

*Figure 9. Matrix generated after algorithm execution*

### iii. Resulting Architecture

The resulting architecture is shown in Figure 10, encompassing the analysis and allocation of the lowest level of the logical component and function. While the current approach proposes a single logical architecture, in case of multiple configurations with the same minimal coupling value exist, one of the relevant solutions is selected by our tool. We may propose evolution to generate various architecture alternatives and support trade-off analyses in future work.

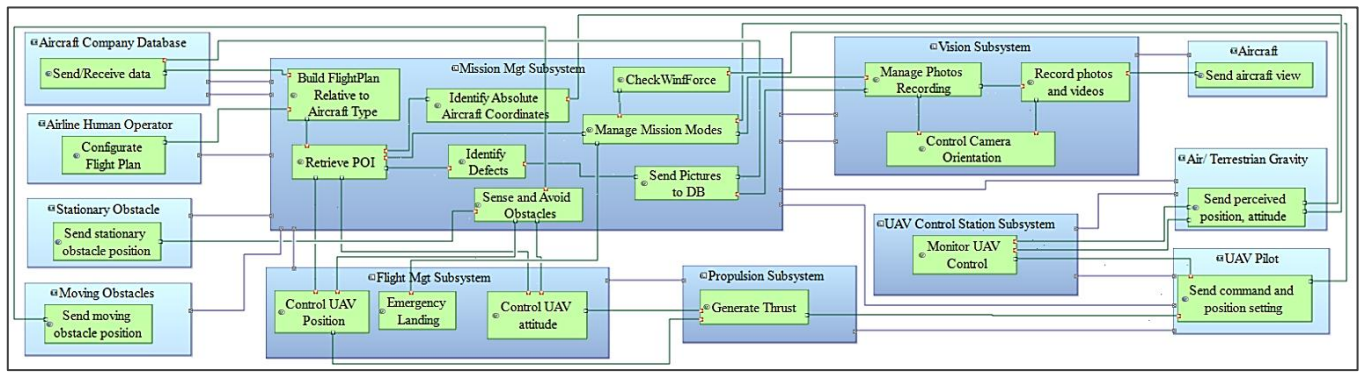

*Figure 10. Allocation result after algorithm execution*

## iv. Genetic Algorithm Parameters Tuning

The setting of the genetic algorithm parameters has been performed currently by iterative experiments. In Table 1., we highlight some experiments we made on AIDA simplified model and a larger model. The genetic parameters have been set accordingly to obtain some viable solutions and keep a reasonable execution time based on the previous work [9]. The approach has been also executed over another Capella model, AIDA large model with a larger parameter.

The genetic algorithm parameters can be modified in the Capella tool via user interface. During the experiment, it was observed that the population size increase with the matrix size, for matrix 23x23 from 200 population size give a coupling value minimized, the survivor percentage have not significant impact less than 90%, else can stay in the loop, the population mutation and the gene mutation have similar impact, the population mutation until 50% and gene mutation <20% centralize function allocation in few logical components and high percentage (>40%) can be out of solution space remaining in the loop.

| Model Name | # Functions | # Components | #Pre-Allocated Functions | Execution time (seconds) | GA Parameters |
|---|---|---|---|---|---|
| AIDA light model | 23 | 12 | 15 | 1,45 | • Initial population=1000<br>• Maximum generation=50<br>• Survivor (crossover) percentage=70%<br>• Parent percentage=20%<br>• Population mutation percentage=70%<br>• Gene mutation percentage=30% |
| AIDA large model | 47 | 17 | 7 | 120 | • Initial population=2500<br>• Maximum generation=200<br>• Survivor (crossover) percentage=70%<br>• Parent percentage=30%<br>• Population mutation percentage=40%<br>• Gene mutation percentage=70% |

*Table 1. Execution time of the algorithm and genetic algorithm parameters*

## v. Comparison to our initial implementation in Cameo

The Table 2. shows comparison results between Capella and Cameo tool for the similar model scope. We can notice that coupling value is larger with Capella model in regards to Cameo SysML model.

| Model name | Coupling value | # Interactions |
|---|---|---|
| AIDA (Cameo tool) | 3.0 | 23 |
| AIDA light model (Capella) | 8.7 | 21 |

*Table 2. Comparative values in different tool implementation*

This can be explained for different reasons. In our initial implementation in the Cameo tool, the model has not pre-allocated functions to external actors, which means they are not included in the analysis. In the other hand, in Capella tool implementation, the external actors have pre-allocated functions, and they are included in the analysis, leading to discrepancies in the coupling values. External actors are considered as constraint for the algorithm in the analysis process, but they are not available for function allocation. The functions available for allocation can only be allocated to the logical components. The number of transitions refer to the interaction between functions allocated in different components of the system.

## IV. Current Status, Future work, and Perspectives

After reviewing with different parties, the usage of such tool may bring interest when there is still flexibility to distribute functions over the systems, which might be difficult when reusing legacy products or when system architecture is highly constrained by organizational aspects. The generic algorithm is also able to consider already allocated functions as constraints for the space exploration.

Our contribution is currently in the initial stage to give the ability to systems engineer to find a ready to use in Capella MBSE use case. As future work we have identified several possibilities to extend this proposal to:

- Provide implementation of DSM generation within SysML Tools as open-source solution
- Extend the concept to introduce consideration of timing constraints and extend the functions and component exchanges with a time delay property and ensure as a constraint that Time budget allocated to the overall functional chain are fulfilled
- Explore other algorithms than Genetic Algorithm and use optimization techniques proposed in related works to handle large matrices.
- Explore the possibility to generate alternatives of architectures in a same model and exhibit the associated properties (timing, performance, costs, …) of each. Then extend this with multi-dimensional optimization techniques.

## V. Related Works

DSM/N2 matrices are known tools to analyse the architecture complexity of systems and there are existing work which extract DSM/N2 matrices from MBSE methods & tools. A first proposal in [11] exhibits how to extract N2 matrix from model of processes developed using OPM method [12]. In [13], the authors propose a contribution compatible with different MBSE tools using the standardized XMI format. Using DSM and genetic algorithm has been investigated in [14]. In this paper, the authors propose a genetic algorithm tailored for problems characterized by modularity, hierarchy, and overlap within complex systems. In [15], the authors proposed to modify DSM's clustering algorithms to include several design constraints. Robert et al. [16] introduce an approach to DSM clustering, addressing shortcomings in existing clustering techniques by encompassing data acquisition and handling multiple perspectives to a post-processing phase that corrects results for technical feasibility. Finally, in [17], authors proposed and optimized fast clustering algorithm using genetic algorithm to generate DSM matrices for complex system architectures. However, this approach is performed on physical architecture level and is presented as a MATLAB script to support optimization of complex multi-physical systems architecture (with a high number of physical interfaces).

## VI. Conclusion

The specificity of our approach is to propose an optimization algorithm of DSM where the clustering algorithm is influenced by the function's allocation to Logical Architecture and is implemented as a genetic algorithm integrated within common MBSE tools (Cameo Systems Modeler, Capella). This allows to address system representation with intermediate complexity using the Logical Architecture concept proposed by several MBSE methodologies such as ARCADIA. As an achievement, we have published our proposal as a Capella addon named **DSM4Capella** in Capella Community within Labs4Capella [18].